\documentclass{article}
\usepackage{graphicx}  
\usepackage{amsmath}   
\usepackage{cite}
\usepackage{amssymb}   
\usepackage{bm} 
\usepackage{dcolumn}
\usepackage{color}
\usepackage{mathrsfs}
\usepackage{amsfonts}
\usepackage{varioref}
\RequirePackage[colorlinks,citecolor=blue,urlcolor=magenta,linkcolor=blue]{hyperref}
\usepackage[colorlinks,citecolor=blue,urlcolor=magenta,linkcolor=blue]{hyperref}
\newcommand{\sech}{~\textrm{sech~}}
\labelformat{section}{Section #1} 
\labelformat{subsection}{Section #1} 
\labelformat{subsubsection}{Section #1}
\labelformat{subsubsubsection}{Section #1}
\labelformat{equation}{Eq.~(#1)} 
\labelformat{figure}{Fig.~#1} 
\labelformat{subfigure}{Fig.~\thefigure#1} 
\labelformat{table}{Tab.~#1} 
\labelformat{appendix}{Appendix #1}
\begin{document}

\title{A comment on generalized Schwinger effect}

\author{Karthik Rajeev\footnote{karthik@iucaa.in} $^{1}$, Sumanta Chakraborty\footnote{sumantac.physics@gmail.com} $^{2}$ and T. Padmanabhan\footnote{paddy@iucaa.in} $^{1}$\\
$^{1}$\small{IUCAA, Post Bag 4, Ganeshkhind, Pune University Campus, Pune 411007, India}\\
$^{2}$\small{Department of Theoretical Physics, Indian Association for the Cultivation of Science, Kolkata-700032, India}}
\maketitle
\begin{abstract}
A spatially homogeneous, time-dependent, electric field can produce charged particle pairs from the vacuum. When the electric field is constant, the mean number of pairs which are produced depends on the electric field and the coupling constant in a non-analytic manner, showing that this result cannot be obtained from the standard perturbation theory of quantum electrodynamics. When the electric field varies with time and vanishes asymptotically, the result may depend on the coupling constant either analytically or non-analytically. We investigate the nature of this dependence  in several specific contexts. We show that the dependence of particle production on coupling constant is non-analytic for a  class of time-dependent electric fields, with the leading order non-analytic behaviour being controlled by a specific parameter which can be identified. We also demonstrate that, for another class of electric fields, which vary rapidly, the dependence of particle production on coupling constant is analytic. Finally, we  describe what happens to these results when we go beyond the leading order, using some specific examples.
\end{abstract}
	\section{Introduction}
	Production of particles by a non-trivial classical background source is an ubiquitous effect in quantum field theory (QFT). Hawking radiation from a black-hole, the production of particles in an expanding universe and the Schwinger effect are some of the most studied examples of this phenomenon \cite{hawking1975particle,Chakraborty:2017pmn,Lochan:2016nbs,Chakraborty:2015nwa,Singh:2014paa,PhysRevLett.21.562,mukhanov_winitzki_2007,calzetta2008nonequilibrium,birrell1982quantum,fulling1989aspects,schwinger1951gauge,Sauter1931,Heisenberg1936,weisskopf1936electrodynamics,kim2016complex}. Even though the physical mechanism of particle creation depends on the specific system we are considering, there are certain common features. For instance, when the classical background is spatially homogeneous, the analysis of particle production in many interesting scenarios reduces to the study of a time-dependent harmonic oscillator. Because of this similarity in the mathematical description, from the study of a specific QFT 
system, one can gain important insights about several other systems (for example, see \cite{Martin:2007bw}). In this paper, we will focus on the mechanism by which charged particle pairs can be produced from vacuum in a spatially homogeneous, time-dependent electric field \cite{greiner1985quantum,Grib:330376,popov1972,itzykson2006quantum,Dunne:2004nc,PhysRevD.75.045013,PhysRevD.73.065020,PhysRevLett.67.2427,PhysRevD.45.2044,Kluger:1992md}, which could be called the generalized Schwinger effect.
	
Besides its mathematical simplicity, there are several features of the Schwinger effect that makes it a promising model for understanding various aspects of QFT. An important such feature is its non-perturbative nature. The rate of particle production in a homogeneous time-independent electric field background depends on the field strength and coupling constant in a non-analytic manner (see \cite{kim2016complex} for a recent review). This shows that one cannot obtain this result from the standard perturbation theory of quantum electrodynamics (QED). While the predictions of perturbative approaches in QFT has been verified to very high precisions \cite{PhysRevLett.92.161802,PhysRevLett.97.030801,PhysRevLett.99.039902,gabrielse2006new}, the verification of non-perturbative results remains as a major challenge in experimental physics. Schwinger effect provides us such an opportunity to understand this relatively unexplored regime of QFT in a better (and possibly deeper) way. Therefore, recently there have been several studies regarding the experimental verification of Schwinger effect  \cite{RevModPhys.84.1177,Dunne2009,Sokolov2014,tuchin2013particle}. Even though the strength of the electric field that is required to test this phenomenon is beyond the current state-of-the-art, it will be accessible to some of the upcoming laser facilities like Extreme Light Infrastructure (ELI) \cite{ELI} and European X-ray Free-Electron Laser (European XFEL) \cite{EXFEL}.
	
While the Schwinger mechanism is an exactly solvable problem, it is practically impossible to realize a spatially homogeneous, constant electric field in an experimental setup. In a laboratory setting, one can only produce electric fields that vanish asymptotically in time though it could be spatially homogeneous to a necessary level of approximation. A well studied configuration of electric field with this property in which the pair production can be analytically computed is the so-called Sauter type electric field given by $\mathbf{E}=(E_0\sech^2(\omega t),0,0)$, where $\omega$ is a dimension-full constant \cite{sauter2,nikishov1,ambjorn,balantekin,PhysRevD.73.065028,PhysRevD.65.105002,PhysRevD.53.7162,PhysRevD.78.105013}. The study of pair production in the Sauter type field reveals an interesting feature of the generalized Schwinger effect. It turns out that the mean number of particles $n_{\mathbf{k}}$, produced  with a given momentum $\mathbf{k}$ in the presence of a Sauter type field can depend on the coupling constant $q$ and field strength $E_0$ either analytically or non-analytically depending on whether the field varies rapidly or adiabatically in time, respectively. In particular, the expression for $n_{\mathbf{k}}$ when the field is adiabatically varying can be shown to approach the corresponding expression for the Schwinger case. Therefore, one expects that the mean number of particles produced for a generic electric field configuration $\mathbf{E}(t)$ may depend on the coupling constant and field strength either analytically or non-analytically depending on certain conditions of the relevant parameters involved. Motivated by this, we seek for two general classes of electric field configurations such that the mean number of particles produced exhibit, either non-analytic dependence on the coupling constant (and field strength) when a specific condition is satisfied or, exhibit analytic dependence on the coupling constant (and field strength).
	
The paper is organized as follows: In \ref{sauterreview}, we review the instability of the vacuum of QED in the presence of a Sauter type electric field. It turns out that the nature of dependence of the particle content on the coupling constant under Sauter type field depends on a parameter $\gamma\equiv m\omega/(|qE_0|)$ where $m$ and $q$ are the mass and the charge of the field respectively. This parameter measures the degree of adiabaticity of the electric field in its time variation. We demonstrate that the mean number of produced particles $n_{\mathbf{k}}$ as well as the probability of pair creation $\mathcal{P}$ depend analytically on $|qE_0|$, when $\gamma\gg1$ and are non-analytic functions of $|qE_0|$, when $\gamma\ll 1$. In \ref{sauter}, we focus on a class of electric fields of the form $\mathbf{E}=(E_0f(\omega t),0,0)$, where $f(s)$ vanishes in the limit $|s|\rightarrow \infty$. We examine the conditions for the validity of perturbative analysis for this system. Further, we show explicitly that,
 when the aforementioned conditions are met, the mean number of particles produced is an analytic function of $|qE_0|$. For the electric field of the form $\mathbf{E}=(E_0f(\omega t),0,0)$, in \ref{nonperturbative}, we consider the scenario in which perturbative analysis fails. We show explicitly that, in this non-perturbative regime, the mean number of particles produced has a factor which is non-analytic in $|qE_0|$, when the integral of $f(s)$ has a certain asymptotic behaviour. Throughout the paper, we use a system of units in which $c=\hbar=1$. We work with $(+,-,-,-)$ signature for the metric tensor. 
\section{Warm-up: Vacuum instability in Sauter type potential}\label{sauterreview}
	
	Let us rapidly review the Schwinger effect in the Sauter type electric field to identify the two regimes in which the effect is perturbative or non-perturbative. Consider a spatially homogeneous electric field along the $x$ direction, with the time dependence:
	\begin{align}
	E(t)=E_0 \sech^{2}\left(\omega t\right)~.
	\end{align}
	The associated vector potential could be taken as $A^{\mu}=(0,A(t),0,0)$ with
	\begin{align}
	A(t)=-\int E(t)dt=-\frac{E_0}{\omega}\tanh \omega t~.
	\end{align}
	For $\omega \rightarrow \infty$, the electric field would change rapidly. In this case, the particle production rate is known to be analytic in $|qE_0|$. While, for $\omega \rightarrow 0$, we can expand the vector potential $A(t)$ in a Taylor series with the following leading order behaviour: $A(t)\approx -(E_0/\omega)\omega t=-E_0t$, which mimics the standard Schwinger effect. Thus, in this limit, mean number of particles produced is expected to be non-analytic in $|qE_0|$.
	
	In order to study the vacuum instability in the presence of this field, one may compute the vacuum persistence probability  $\mathcal{P}$. It can be shown that, under the weak field approximation, this probability is given by $\mathcal{P}=\exp(-\mathcal{A}_{E})$ where $\mathcal{A}_{E}$ is the Euclidean action evaluated for an instanton solution of the equation of motion $d\mathbf{p}/dt=q\mathbf{E}$, where $\mathbf{p}$ is the momentum \cite{affleck1982pair,Dumlu:2011cc,PhysRevD.65.105002,PhysRevD.73.065028,sbtp}. This procedure works whenever, the solution exhibits periodicity in the Euclidean time. In particular, for the Sauter-type electric field the Euclidean action for a trajectory with $p_y=p_z=0$ evaluated over one period in imaginary time is given by
	\begin{align}\label{Euc_Ac}
	\mathcal{A}_{\rm E}=\frac{\pi m}{\omega}\frac{2\gamma}{1+\sqrt{1+\gamma ^{2}}}~,
	\end{align}
	where, $\gamma=m\omega/|qE_{0}|$. (It may be noted that the Euclidean action corresponding to an instanton of winding number $n$ is given by $n\mathcal{A}_{\textrm{E}}$. However, for simplicity, we will restrict our discussion to the case when $n=1$.) We can compute the probability for vacuum persistence, $\mathcal{P}\approx \exp( -\mathcal{A}_{\rm E})$ in the two limits: (a) $\gamma \ll 1$ and (b) $\gamma \gg 1$. As mentioned earlier, we expect $\mathcal{P}$ to be non-analytic in the field strength in the case $\gamma \ll 1$, (which includes the case of a constant electric field), while it could be analytic  for $\gamma \gg 1$. For $\gamma \gg 1$ case, from \ref{Euc_Ac} it follows that the probability for pair production becomes,
	\begin{align}
	\mathcal{P}=e^{-\frac{2 \pi  m}{\omega }}
	\left(1+\frac{2 \pi  m}{\omega 
		\gamma }+\mathcal{O}(\gamma^{-2})\right)\approx e^{-\frac{2 \pi  m}{\omega}}
	\left(1+\frac{2 \pi qE_0}{\omega^2 
	}\right)~.
	\end{align}
	Thus $\mathcal{P}$ is independent of the field strength to leading order in $1/\gamma$. Hence in this limit the pair production probability is indeed analytic in the field strength. 
	On the other hand, when  $\gamma \ll 1$, the pair production probability takes the following form
	\begin{align}\label{prob_nonpert}
	\mathcal{P}=\exp\left\{-\frac{\pi m^{2}}{qE_0}\left(1-\frac{\gamma ^{2}}{4}\right)+\mathcal{O}\left(\gamma ^{3}\right)\right\}\approx \exp\left\{-\frac{\pi m^{2}}{qE_0}\right\}\left(1+\frac{\pi m^{2}}{qE_0}\frac{m^2\omega ^{2}}{4(qE_0)^2}\right)~.
	\end{align}
	Thus it is clear that for small $\gamma$, the pair production probability is non-analytic, as anticipated. This is mainly due to the fact that $\gamma \ll 1$ corresponds to $\omega \rightarrow 0$ limit and hence is similar to a constant electric field. 
	\subsection{Mean number of particles produced}
	
	The mean number of particles produced in the asymptotic future can be found from the Bogoliubov coefficients connecting the `in-modes' and `out-modes'. Given the spatial homogeneity of the problem, the Fourier transform $\phi_{\mathbf{k}}$ of a complex scalar field $\phi$ of charge $q$ and mass $m$ in the Sauter type electric field introduced in \ref{sauterreview} satisfies the following differential equation 
	\begin{align}
	\ddot{\phi}_\mathbf{k}+\left[m^2+|\mathbf{k}_{\perp}|^2+\left(k_x-\frac{qE_0}{\omega}\tanh \omega t\right)^2\right]\phi_{\mathbf k}=0~.
	\end{align}
	The solutions to this equation which corresponds to the `in' and `out' modes can be written in terms of Hypergeometric functions \cite{PhysRevD.53.7162,PhysRevD.78.105013}. The mean number of `$\mathbf{k}$-mode' particles produced in the asymptotic future $n_{\mathbf{k}}$ can then be found from the Bogoliubov coefficients to be given by\cite{PhysRevD.53.7162,PhysRevD.78.105013}:
	\begin{align}\label{particle}
	n_\mathbf{k}=\frac{\cosh^2\left[\pi\sqrt{\left(\frac{qE_0}{\omega^2}\right)^2-\frac{1}{4}}\right]+\sinh^2\left[\frac{\pi}{2\omega}(\omega_{+}-\omega_-)\right]}{\sinh\left(\frac{\pi\omega_-}{\omega}\right)\sinh\left(\frac{\pi\omega_+}{\omega}\right)}~.
	\end{align}
	where, $\omega_{\pm}=\{m^2+k_y^2+k_z^2+\left(k_x\mp (qE_0)/\omega\right)^2\}^{1/2}$. The two limits discussed in the previous section corresponds  to the following two limits in this particular case: (a) $\omega \ll \sqrt{qE_0}$, which corresponds to $\gamma \ll 1$ in the previous scenario, and (b) $\omega \gg (qE_0/\sqrt{m^{2}+\mathbf{k}^{2}})$, equivalent to $\gamma \gg 1$. Here $\mathbf{k}^{2}=k_{x}^{2}+|\mathbf{k}_{\perp}|^{2}$. When $\omega \ll \sqrt{qE_0}$, it immediately follows that
	\begin{align}\label{particle-sauter}
	n_{\mathbf{k}}=\exp\left(-\frac{\pi(m^2+|\mathbf{k}_{\perp}|^2)}{qE_0}\right)~,
	\end{align}
	which coincides with the pair production probability derived in the previous section associated with $\gamma \ll 1$. As evident and anticipated, the mean number of particles produced is non-analytic in the field strength. The probability $\mathcal{P}(\mathbf{k})$ that a particle with momentum $\mathbf{k}$ and an anti-particle with momentum $-\mathbf{k}$ is produced is given by $\mathcal{P}(\mathbf{k})=n_{\mathbf{k}}/(1+n_\mathbf{k})$. When the electric field is weak (i.e., $qE_0/m\ll 1$), so that $n_\mathbf{k}\ll 1$, the probability of pair production becomes $\mathcal{P}(\mathbf{k})\approx n_{\mathbf{k}}$. Therefore, for $k_z=k_y=0$ and the case of weak electric field, which was the case considered in the previous section, the probability of pair production to leading order in $qE_0$ becomes $\mathcal{P}(\mathbf{k})=\exp(-\pi m^2/qE_0)$, which is in agreement with \ref{prob_nonpert}.
	
	The opposite limit, corresponding to $\omega \gg (qE_0/\sqrt{m^{2}+\mathbf{k}^{2}})$, can also be easily obtained from \ref{particle}, leading to,
	\begin{align}\label{sauter_perturb}
	n_\mathbf{k} \approx\frac{\left(\frac{\pi qE_0k_x}{\omega^2\varepsilon}\right)^2}{\sinh^2\left(\frac{\pi \epsilon}{\omega}\right)}~,
	\end{align}
	where, $\varepsilon^2=\mathbf{k}^2+m^2$. Clearly, $n_{\mathbf{k}}$ is analytic in $qE_0$ in conformity with the result in the previous section. Again, we see that the particle production is non-analytic in the coupling constant for small $\omega$, while is analytic for large $\omega$. In the first case, the mean number of particles produced, to the leading order coincides with that of the standard Schwinger effect.
	\section{Analytic and non-analytic dependences in a general context}\label{sauter}
	
	We will now generalise the previous results --- which were obtained for a specific form of the electric field, viz. the Sauter field --- to a more general configuration.
	For this analysis  we will consider an electric field along the $x$-axis of the form \cite{popov1972}
	\begin{align}\label{defineE}
	E(t)=E_0f(\omega t)~,
	\end{align}
	where, $E_0$ is a constant. The vector potential can then be chosen to be $(0,A,0,0)$, where
	\begin{align}\label{defineA}
	A(t)=-\frac{E_0}{\omega}F(\omega t)~,
	\end{align}
	where $F$ is defined through  $dF(s)/ds=f(s)$. Since any physically realizable electric field must vanish as $t\rightarrow\infty$, we may impose the condition that $f(\omega t)$ vanishes as we approach the asymptotic past and future times. Therefore, the function $F$ satisfies 
	\begin{align}\label{vanish}
	\lim\limits_{|s|\rightarrow\infty}\frac{dF}{ds}=0~.
	\end{align}
	In this external background electric field, the Fourier transform $\phi_{\mathbf{k}}$ of a complex scalar field of mass $m$ and charge $q$ satisfies the following time dependent harmonic oscillator equation:
	\begin{align}\label{eqforphik}
	\ddot{\phi_{\mathbf{k}}}+\omega_{\mathbf{k}}^2\phi_{\mathbf{k}}=0~,
	\end{align}
	where the time dependent frequency $\omega_{\mathbf{k}}$ is given by
	\begin{align}\label{defineomega}
	\omega_\mathbf{k}^2(t)=\varepsilon ^2+\frac{2 k_x m F(\omega 
		t)}{\gamma }+\frac{m^2 F(\omega  t)^2}{\gamma
		^2}~.
	\end{align}
	and $\varepsilon^2=\mathbf{k}^2+m^2$. We will now study this equation and its solutions in the two appropriate limits.
	\subsection{Perturbative limit}\label{pertlimit}
	
	In this subsection we will assume that the function $F$ is bounded as follows:
	\begin{align}\label{cond_01}
	\textrm{Condition 1}: \qquad \qquad F_{\rm min}<F<F_{\rm max}~.
	\end{align} 
	Moreover, from \ref{defineomega}, we see that the field strength dependent part of $\omega_{\mathbf{k}}^2$ can be treated as a perturbation whenever,
	\begin{align}\label{pertcondition}
     \textrm{Condition 2}: \qquad \qquad	F_{-}\ll F_{\textrm{min}} <F(t)<F_{\textrm{max}}\ll F_{+}~,
	\end{align}
	where, $F_{\pm}=\gamma(-k_{x}\pm\sqrt{k_x^2+\varepsilon^2})$ are the roots of the equation $2 k_x m F/\gamma+m^2 F^2/\gamma^2=\varepsilon ^2$.
	When this condition is satisfied, we can solve \ref{eqforphik} perturbatively in powers of $1/\gamma$. Therefore, if an electric field satisfies Condition 1 and 2 respectively, we can take the solution to \ref{eqforphik} to be of the following form
	\begin{align}
	\phi_{\mathbf{k}}=\phi_{\mathbf{k}(0)}+\frac{1}{\gamma}\phi_{\mathbf{k}(1)}+\frac{1}{\gamma^2}\phi_{\mathbf{k}(2)}+\cdots~.
	\end{align} 
	The $\mathcal{O}(1/\gamma)$ solution which approaches a positive frequency mode at $t\rightarrow-\infty$ can be easily found to be
	\begin{align}
	\phi_{\mathbf{k}}=e^{i\varepsilon t}-\frac{2mk_x}{\gamma\varepsilon}\int_{-\infty}^{t}dt'\sin[\varepsilon(t-t')]F(\omega t')e^{i\varepsilon t'}+\mathcal{O}(\gamma^{-2})~.
	\end{align}
	Due to the appearance of $\sin[\varepsilon(t-t')]$, we see that a positive frequency mode in the asymptotic past evolves into a linear combination of both positive and negative frequency modes. That is
	\begin{align}
	\phi_{\mathbf{k}}\approx
	\begin{cases}
	e^{i\varepsilon t},~~t\rightarrow-\infty\\
	\mathcal{A}e^{i\varepsilon t}+\mathcal{B}^*e^{-i\varepsilon t},~~ t\rightarrow \infty
	\end{cases}
	\end{align} 
	where, $\mathcal{A}=1+\mathcal{O}(1/\gamma)$ and 
	\begin{align}
	\mathcal{B}^*=\left(\frac{\pi k_xqE_0}{i\omega\varepsilon^2}\right)\tilde{f}\left(\frac{2\varepsilon}{\omega}\right)+\mathcal{O}(\gamma^{-2})~,
	\end{align}
	are the associated Bogoliubov coefficients. Furthermore, $\tilde{f}(s)$ is the Fourier transform of $f(\tau)$. Therefore, the mean number of particles produced is given by (see also \cite{popov1972})
	\begin{align}\label{perturb}
	n_{\mathbf{k}}(\infty)=\frac{|\pi k_x(qE_0)\tilde{f}\left(\frac{2\epsilon}{\omega}\right)|^2}{\omega^2\varepsilon^4}+\mathcal{O}(\gamma^{-4})~.
	\end{align}
	Note that this expression is a Taylor series in $qE_0$ and hence analytic in the same. Therefore, we have shown in this section that: \textit{if the electric field of the form given by \ref{defineE} satisfies Conditions 1 and Condition 2, the asymptotic value of the mean number of particles produced $n_{\mathbf{k}}$ is an analytic function of $|qE_0|$.}
	
	As a consistency check, let us apply this to the case of Sauter type electric field that we discussed in \ref{sauter}. In this case the relevant $f(\omega t)$ is given by $\sech^2(\omega t)$, whose Fourier transform evaluated at $2\varepsilon/\omega$ turns out to be $\omega^{-1}\varepsilon~ \textrm{csch}(\pi\omega/\varepsilon)$. Therefore, from \ref{perturb}, the mean number of particles produced to leading order in $\gamma$ is given by
	\begin{align}
	n_\mathbf{k}=\frac{\left(\frac{\pi qE_0k_x}{\omega^2\varepsilon}\right)^2}{\sinh^2\left(\frac{\pi}{\omega}\sqrt{\mathbf{k}^2+m^2}\right)}+\mathcal{O}(\gamma^{-4})~,
	\end{align}  
	which is in complete agreement with \ref{sauter_perturb}.
	\subsection{Non-Perturbative Limit}\label{nonperturbative}
	
	In this subsection we will consider electric field configurations of the form \ref{defineE}, which violates Condition 1 and 2, presented in \ref{cond_01} and \ref{pertcondition} respectively. This corresponds to the case in which the field dependent part of $\omega_{\mathbf{k}}^2$ is much larger than $\varepsilon^2$ in some range of time. We focus on the situation which satisfies the following conditions: 
	\begin{align}\label{cond_03}
	\textrm{Condition 3:}\qquad |F(t)|\gg F_{+},|F_{-}|~~\textrm{when}~~|t|\gg t_c~,
	\end{align}
	where $t_c$ is some critical time. This implies that the perturbation theory cannot be used to find a solution to \ref{eqforphik} valid for all times $t$. However, since we have assumed the field to vanish asymptotically, from \ref{vanish} we see that the WKB approximation becomes valid at late and early times. The simplest example of this case is given by a constant field. To set the stage, we will briefly review this special case before doing a general analysis.

	The Fourier modes $\phi_{\mathbf{k}}$ of a complex scalar field in a constant electric field is given by $\mathbf{E}=(E_0,0,0)$, which satisfies the following harmonic oscillator equation, such that,
	\begin{align}\label{schwinger_oscillator}
	\ddot{\phi}_{\mathbf{k}}+\omega_{\mathbf{k}}^2(t)\phi_{\mathbf{k}}=0~
	\end{align}
	where, for the choice of gauge $A_{\mu}=(0,-E_{0}t,0,0)$ one have,
	\begin{align}\label{freq_schwinger}
	\omega_{\mathbf{k}}^2(t)=m^2+|\mathbf{k}_{\perp}|^2+(qE_0t+k_x)^2~.
	\end{align}
	The exact solutions of \ref{schwinger_oscillator} are known in terms of parabolic cylinder functions. But, the exact number of particles produced at $t=+\infty$ can also be found using the WKB solutions of \ref{schwinger_oscillator} in the limit $|t|\rightarrow\infty$.
	Let us denote by $\xi_{\mathbf{k}(\textrm{in})}$, the `in-modes', which are solutions to \ref{schwinger_oscillator} that behave as positive frequency modes at $t\rightarrow-\infty$. Similarity, we define the `out-modes' $\xi_{\mathbf{k}(\textrm{out})}$, which are the positive frequency solutions of \ref{schwinger_oscillator} at $t\rightarrow\infty$. From \ref{freq_schwinger}, we see that $(|\dot{\omega}_{\mathbf{k}}|/\omega_{\mathbf{k}}^2)\ll1$ as $|t|\rightarrow\infty$. Hence, in the asymptotic past and future the WKB approximation is valid. In these regions, the approximate positive frequency solutions $\xi_{\textrm{in}/\textrm{out}}$ behave as $(|\omega_{\mathbf{k}}|)^{-1/2}\exp\left(i\int|\omega_{\mathbf{k}}|dt\right)$. In order to explicitly calculate this WKB limit, let us first look at the behaviour of $|\omega_{\mathbf{k}}|$ as $|t|\rightarrow \infty$, leading to,
	\begin{align}\label{omegaseries}
		|\omega_{\mathbf{k}}(t)|=\begin{cases}
		-qE_0t-\frac{\mathbf{k}_{\perp}^2+m^2}{2
			qE_0
			t}+\mathcal{O}\left(\left\{\frac{1}{qE_0t}\right\}^3\right),~~~t\rightarrow-\infty,\\
				qE_0t+\frac{\mathbf{k}_{\perp}^2+m^2}{2
			qE_0
			t}+\mathcal{O}\left(\left\{\frac{1}{qE_0t}\right\}^3\right),~~~t\rightarrow\infty.
		\end{cases}
	\end{align}
After a straightforward integration, the exponential term, namely $\exp\left(i\int|\omega_{\mathbf{k}}|dt\right)$ simplifies to
\begin{align}
	\exp\left(i\int|\omega_{\mathbf{k}}|dt\right)\approx\begin{cases}
	\exp\left\{-\frac{iqE_0t^2}{2}-\frac{i\lambda}{2}\log(-\sqrt{qE_0}t)\right\}, t\rightarrow-\infty\\
	\exp\left\{\frac{iqE_0t^2}{2}+\frac{i\lambda}{2}\log(\sqrt{qE_0}t)\right\}, t\rightarrow \infty.
	\end{cases}
\end{align} 	
where, $\lambda=(|\mathbf{k}_{\perp}|^2+m^2)/(qE_0)$. Therefore, the WKB approximations for the `in' and `out' modes become	
	\begin{align}\label{xi_in}
	\textrm{As $t\rightarrow-\infty$:~~}&\xi_{\mathbf{k}(\textrm{in})} \approx~(-\sqrt{qE_0}t)^{-i\lambda/2-1/2}\exp\left(-\frac{iqE_0t^2}{2}\right)~,
	\\\label{xi_out}
	\textrm{As $t\rightarrow\infty$:~~}&\xi_{\mathbf{k}(\textrm{out})}\approx~(\sqrt{qE_0}t)^{i\lambda/2-1/2}\exp\left(\frac{iqE_0t^2}{2}\right)~,
	\end{align}
	 Let us consider the evolution of $\xi^*_{\mathbf{k}(\textrm{in})}$ from $t\rightarrow-\infty$ to $t\rightarrow\infty$. Since $\{\xi_{\mathbf{k}(\textrm{out})},\xi_{\mathbf{k}(\textrm{out})}^{*}\}$ is a complete set of solutions of \ref{schwinger_oscillator}, we can write $\xi _{\mathbf{k}(\textrm{in})}$ as a linear combination of $\xi_{\mathbf{k}(\textrm{out})}$ and $\xi_{\mathbf{k}(\textrm{out})}^{*}$, such that
	\begin{align}\label{in_is_out_plus_out}
	\xi_{\mathbf{k}(\textrm{in})}^*=\mathcal{A}\xi_{\mathbf{k}(\textrm{out})}^*+\mathcal{B^*}\xi_{\mathbf{k}(\textrm{out})}
	\end{align} 
	where, $\mathcal{A}$ and $\mathcal{B}$ are the standard Bogoliubov coefficients. The 
	mean number of particles produced $n_{\mathbf{k}}$ is then given by $|\mathcal{B}|^2$. One can use the asymptotic expansions  of the parabolic cylinder functions to compute $\mathcal{B}$. But there is simpler and more elegant procedure  (for example, see \cite{landau2013quantum,Padmanabhan:1991uk,Srinivasan:1998ty}), which can be used to obtain $\mathcal{B}$ and hence, $n_{\mathbf{k}}$. To follow this procedure, we start with the WKB approximation for $\xi_{\mathbf{k}(\textrm{in})}^*$, which using \ref{xi_in}, \ref{xi_out} and \ref{in_is_out_plus_out} is given by
	\begin{align}
	\xi_{\mathbf{k}(\textrm{in})}^*=
	\begin{cases}
	(-\sqrt{qE_0}t)^{i\lambda/2-1/2}\exp\left(\frac{iqE_0t^2}{2}\right),&~~t\rightarrow-\infty\\
	\mathcal{A}(\sqrt{qE_0}t)^{-i\lambda/2-1/2}\exp\left(-\frac{iqE_0t^2}{2}\right)+\mathcal{B}^*(\sqrt{qE_0}t)^{i\lambda/2-1/2}\exp\left(\frac{iqE_0t^2}{2}\right)~.&~~t\rightarrow\infty
	\end{cases}
	\end{align}
	Now, in the asymptotic expression for $\xi_{\mathbf{k}(\textrm{in})}^*$ as $t\rightarrow-\infty$, if we treat $t$ as a complex variable and rotate $t$ in the complex plane from $\textrm{arg}[t]=0$ to $\textrm{arg}[t]=\pi$, then: (i) the exponential part $\exp\left[(iqE_0t^2)/2\right]$ transforms to itself and (ii) the pre-factor, namely $(-\sqrt{qE_0}t)^{i\lambda/2-1/2}$), transforms to that of $\xi_{\textrm{out}}$ times $e^{-i\pi/2}e^{-\pi\lambda/2}$. Hence we see that  the asymptotic
	expression for $\xi^*_{\textrm{in}}$, under the rotation of $t$ in the complex plane from $\textrm{arg}[t]=0$ to $\textrm{arg}[t]=\pi$, nicely gets mapped to the asymptotic
	expression for $\xi_{\textrm{out}}$ as $t\rightarrow-\infty$. Therefore, we read off the Bogoliubov coefficient $\mathcal{B}$ to be:
	\begin{align}
	\mathcal{B}=e^{-i\pi/2}e^{-\pi\lambda/2}~.
	\end{align}
	Hence, the mean number of particles produced is given by $n_{\mathbf{k}}=|\mathcal{B}|^{2}=e^{-\pi\lambda}$, which is the standard result. 
	
	To generalise this result we would like to point out that the above result can also be derived from a different perspective. In particular, we note that using \ref{omegaseries} it is possible to write down the time dependent frequency $\omega_{\mathbf{k}}$ having the following series expansion near $|t|\rightarrow\infty:$
	\begin{align}\label{omegafornonanalyticity}
	\omega_{\mathbf{k}}(t)&\approx \kappa|t|\left(1+\frac{\omega_0^2}{2\kappa^2|t|^2}+...\right)=\sum_{n=-\infty}^{1}C_n(\kappa)|t|^{2n-1}~,
	\end{align}
	where, $\kappa \equiv(qE_0)$ and $\omega_0^2\equiv(m^2+|\mathbf{k}_{\perp}|^2)$. Note that there are only odd powers of $|t|$ present in the above expansion of $\omega_{\mathbf{k}}(t)$ near $|t|=\infty$. We observe that the mean number of particles produced, found from rotating $t$ in the complex plane, is related to $C_0$ as
	\begin{align}
	n_{\mathbf{k}}=e^{-2\pi C_0}~,
	\end{align} 
	and the non analyticity comes from the fact that $C_0\propto 1/(qE_0)$. We will now show that this feature continues to hold even in a more general context.
	
	Motivated by the above observation, let us consider a class of electric fields satisfying Condition 3 as in \ref{cond_03}, for which the mean number of particles produced is non-analytic in the coupling constant. This class is characterized by the following properties:
	\begin{itemize}
		
		\item For, $|\tau|\rightarrow\infty$, $f(\tau)$ is symmetric under $\tau\rightarrow-\tau$.
		
		\item $F(\tau)$ diverges as a power series, as $|\tau|\rightarrow\infty$.
		\end{itemize}
	where, the functions $f$ and $F$ are defined in \ref{defineE} and \ref{defineA}, respectively. These two conditions imply that $F(|\tau|)$ has the following expansion near infinity
	\begin{align}\label{Asymp_F}
	F(|\tau|)\approx\sum_{n=-\infty}^{\infty}\mathcal{C}_n|\tau|^{2n-1}~.
	\end{align} 
	For later use, let us also note the asymptotic series for $1/F(|\tau|)$,
	\begin{align}\label{Asymp_iF}
	\frac{1}{F(|\tau|)}=\sum_{n=-\infty}^{\infty}\tilde{\mathcal{C}}_n|\tau|^{2n-1}~,
	\end{align}
	where, the coefficients $\mathcal{C}_n$ and $\tilde{\mathcal{C}}_n$ satisfy,
	\begin{align}
	\sum_{n=-\infty}^{\infty}\mathcal{C}_n\tilde{\mathcal{C}}_{l-n}=
	\begin{cases}
	1,~~l=1\\
	0,~~\forall l\in\mathbb{Z}-\{1\}
	\end{cases}
	\end{align}
	where $\mathbb{Z}$ stands for the set of all non-negative integers. Recall that the Fourier modes $\phi_{\mathbf{k}}$ of a complex scalar field in this kind of electric field satisfies the equation of motion of a harmonic oscillator of unit mass and frequency $\omega_{\mathbf{k}}$ given by \ref{defineomega}. Since $F(\omega t)\gg 1$ as $|t|\rightarrow\infty$, $\omega_\mathbf{k}$ has the following approximation in this limit.
	\begin{align}
	\omega_\mathbf{k}(t)\approx \frac{m F(\omega  t)}{\gamma }+k_x +\frac{\gamma  (|\mathbf{k}_{\perp}|^2+m^2)}{2 m F(\omega t)}+\mathcal{O}\left(F(\omega t)^{-2}\right)~.
	\end{align}
	Proceeding exactly as in  our analysis of the standard Schwinger effect, the WKB approximation for the `in' and `out' modes can be found to be
	\begin{align}
	\textrm{As $t\rightarrow-\infty$:~~}\xi_{\mathbf{k}(\textrm{in})}\approx ~(|qE_0F(\omega t)|)^{-1/2}&\exp\Big\{i\frac{qE_0}{\omega}\int_{-t_0}^{t} dt' F(\omega t')+ik_x(t+t_0)
	\nonumber
	\\
	&+\frac{i\gamma  (\mathbf{k}_{\perp}^2+m^2)}{2 m }\int_{-t_0}^{t}\frac{dt'}{F(\omega t')}\Big\}\\\label{xiout}
	\textrm{As $t\rightarrow\infty$:~~}\xi_{\mathbf{k}(\textrm{out})}\approx ~(|qE_0F(\omega t)|)^{-1/2}&\exp\Big\{-i\frac{qE_0}{\omega}\int_{t_0}^{t} dt' F(\omega t')-ik_x(t-t_0)
	\nonumber
	\\
	&-\frac{i\gamma  (\mathbf{k}_{\perp}^2+m^2)}{2 m }\int_{t_0}^{t}\frac{dt'}{F(\omega t')}\Big\}~.
	\end{align}
	where, for convenience, we have chosen a reference time $t_0$ such that $|t|\gg|t_0|>|t_c|$. Given the power series expansions for $F(\tau)$ and $1/F(\tau)$ a close look at the two integral expressions in the exponential factor of $\xi_{\textrm{in}}$, as $t\rightarrow-\infty$, reveals, 
	\begin{align}
	\int_{-t_0}^{t} dt' F(\omega t')\approx \sum_{n\neq0}\left(\frac{\omega^{2n-1}\mathcal{C}_n}{2n}t^{2n}\right)+\frac{\mathcal{C}_0}{\omega}\log(-\omega t)\\
	\int_{-t_0}^{t} \frac{dt'}{F(\omega t')}\approx \sum_{n\neq0}\left(\frac{\omega^{2n-1}\tilde{\mathcal{C}}_n}{2n}t^{2n}\right)+\frac{\tilde{\mathcal{C}}_0}{\omega}\log(-\omega t)~.
	\end{align}
	Therefore, the asymptotic expression for $\xi_{\textrm{in}}$ simplifies to
	\begin{align}\label{xi_in_gen}
		\xi_{\mathbf{k}(\textrm{in})}\approx ~(|qE_0F(\omega t)|)^{-1/2}(-\omega t)^{i\left(\bar{\lambda}\mathcal{C}_0+\frac{\lambda\tilde{\mathcal{C}}_0}{2}\right)}&\exp\Big\{i\frac{qE_0}{\omega}\sum_{n\neq0}\left(\frac{\omega^{2n-1}\mathcal{C}_n}{2n}t^{2n}\right)+ik_x(t)
		\nonumber
		\\
		&+\frac{i\gamma  (\mathbf{k}_{\perp}^2+m^2)}{2 m }\sum_{n\neq0}\left(\frac{\omega^{2n-1}\tilde{\mathcal{C}}_n}{2n}t^{2n}\right)\Big\}
	\end{align}
where, $\bar{\lambda}=(qE_0)/(\omega^2)$. Similarly, we can show that the asymptotic expression for $\xi_{\textrm{out}}$ is given by
\begin{align}\label{xi_out_gen}
	\xi_{\mathbf{k}(\textrm{out})}\approx ~(|qE_0F(\omega t)|)^{-1/2}(-\omega t)^{-i\left(\bar{\lambda}\mathcal{C}_0+\frac{\lambda\tilde{\mathcal{C}}_0}{2}\right)}&\exp\Big\{-i\frac{qE_0}{\omega}\sum_{n\neq0}\left(\frac{\omega^{2n-1}\mathcal{C}_n}{2n}t^{2n}\right)-ik_x(t)
	\nonumber
	\\
	&-\frac{i\gamma  (\mathbf{k}_{\perp}^2+m^2)}{2 m }\sum_{n\neq0}\left(\frac{\omega^{2n-1}\tilde{\mathcal{C}}_n}{2n}t^{2n}\right)\Big\}
\end{align}
Again, since $\xi_{\mathbf{k}(\textrm{in})}^*$ can be written as a linear combination of $\xi_{\mathbf{k}(\textrm{out})}$ and $\xi_{\mathbf{k}(\textrm{out})}^*$ we have,
	\begin{align}
	\xi_{\mathbf{k}(\textrm{in})}^*=
	\mathcal{A}\xi_{\mathbf{k}(\textrm{out})}^*+\mathcal{B^*}\xi_{\mathbf{k}(\textrm{out})}
	\end{align}
	where, for the relevant asymptotic behaviour one has to use \ref{xi_in_gen} and \ref{xi_out_gen} respectively. Let us now treat $t$ as a complex variable in the expression for $\xi_{in}^*$ obtained from complex conjugation of \ref{xi_in_gen} and rotate $t$ in the complex plane from $\textrm{arg}[t]=0$ to $\textrm{arg}[t]=\pi$. As in the previous case, under the rotation: (i) the exponential factor in the expression for $\xi_{in}^*$ near $t\rightarrow-\infty$ maps to the exponential factor for the expression for $\xi_{\mathbf{k}(\textrm{out})}$ near $t\rightarrow\infty$ and (ii) the pre-factor in $\xi_{\textrm{in}}^*$ maps to that of $\xi_{\textrm{out}}$ except for a constant factor, which is interpreted as the Bogoliubov coefficient $\mathcal{B}$, to be given by
	\begin{align}
	\mathcal{B}=e^{-i\pi/2}e^{-\pi\lambda\tilde{\mathcal{C}}_0/2}e^{-\pi\bar{\lambda}\mathcal{C}_0}~,
	\end{align}
	where we have retained the notation $\lambda=(|\mathbf{k}_{\perp}|^2+m^2)/(qE_0)$ that we used for the case of a constant electric field. Therefore, we see that the mean number of particles produced in has the following \textit{non perturbative} part
	\begin{align}\label{non-perturb-final}
	n_{\mathbf{k}(\textrm{non-pert})}=\exp\Big\{-\frac{\pi(m^2+|\mathbf{k}_{\perp}|^2)\tilde{\mathcal{C}}_0}{qE_0}\Big\}~.
	\end{align}
	Thus we have explicitly demonstrated that: For the class of electric fields with the asymptotic properties mentioned above, \emph{with $\tilde{C_0}\neq0$,} the asymptotic value of the number of particle produced, $n_\mathbf{k}$, will have \emph{a non-analytic dependence on the coupling constant} given by Eq.(48). It is amusing to note that this leading non-analytic behaviour is completely controlled by the coefficient $\tilde{C_0}$ \emph{when it is non-zero}. 
	
	As another example to this result and as a consistency check, consider the asymptotic expansion of the vector potential for the Sauter type potential in the limit $\omega\ll 1/\sqrt{qE_0}$. The corresponding expansion of $1/F$ for this case becomes, $1/F(|\tau|)\sim 1/|\tau|$, implying that, $\tilde{\mathcal{C}_{0}}=1$. So the particle number, following 
	\ref{non-perturb-final} is given by
	\begin{align}
	n_{\mathbf{k}(\textrm{non-pert})}\approx \exp\Big\{-\frac{\pi(m^2+|\mathbf{k}_{\perp}|^2)}{qE_0}\Big\}~,
	\end{align}
	which reproduces the result for the mean number of particles produced in the case of Sauter potential as in \ref{particle-sauter}. However the results presented above emerged from the leading order computation, we will now try to understand what happens to the particle number, in particular to the non-analyticity if we go beyond the leading order. 
\subsection{Beyond Leading Order}\label{beyond_lead}
	
We have explicitly demonstrated in the previous sections that whenever $F(\tau)$ and its inverse has an asymptotic expansion of the form as in \ref{Asymp_F} and \ref{Asymp_iF} respectively with a non-zero $\tilde{C}_{0}$, the particle number will be non-analytic and controlled essentially by $\tilde{C}_{0}$. \textit{While this is a significant result by itself},  it will be interesting to investigate what happens in more general situations. In particular, we can ask what happens when $\tilde{C}_{0}$ vanishes? When we take next-to-leading order terms into account do we get sub-leading non-analytic contributions or does the result become analytic? We will address different aspects of this question in this section. This will not only shed light on the analytic/non-analytic behaviour of particle numbers beyond the leading order but will also provide a new framework to answer the above intriguing questions in a rather counter-intuitive manner.
	
When $\tilde{C}_{0}$ is non-zero, we could obtain the result in a rather \emph{elegant manner} using the analysis in the complex plane (which we will call ``the Landau approach''). Unfortunately, like many elegant tricks, the Landau approach does \emph{not} work when $\tilde{C}_{0}$ vanishes. This failure is due to a subtle technical reason, related  to the non-existence of a logarithmic term in the frequency integral. Since the approach fails, we have delegated the discussion of the reason for its failure to  \ref{App_Add}. 

This implies that one cannot hope for general, elegant, results when we proceed beyond the leading order unless we choose the class of electric fields judiciously. It turns out that, 
despite the failure of the Landau approach, it is indeed possible to address the relevant question in a satisfactory manner for those electric field configurations which allow an analysis based on the Euclidean action method. 
This method essentially involves computing the Euclidean action after the analytic continuation $t \rightarrow it_{E}$ has been performed. Of course, for a \textit{general} potential such an analytic continuation will not lead to sensible results. So for our purpose of investigating the next-to-leading approximation, we will confine ourselves to a  well-defined subclass of potentials which will exhibit the appropriate transition from analytic to non-analytic behaviour. 

In particular we will assume the following conditions to hold --- (a) The factor $F(\omega t)$ appearing in the vector potentials must satisfy, $F^{2}(i\omega t_{\rm E})=-F^{2}(\omega t_{\rm E})$; (b) For finiteness of Euclidean action, we will restrict the range of the integration over Euclidean time $t_E$ to be finite. In what follows we will content ourselves with the vector potentials which will satisfy the above conditions. 

Given the fact that $t_E$ can vary over finite range, we can scale the potential such that this range is confined to the condition $-1<F(\omega t_{\rm E})/\gamma<1$. (We will see explicit examples later on of such cases.) If we further introduce new variable $u$, such that, $F(\omega t_{\rm E})=\gamma u$, then we will restrict $u\in (-1,1)$ and use the integration variable  $du=(\omega/\gamma)F'dt_{\rm E}$. (Aside: When the potential has functions like e.g., $\cosh\omega t$ etc, the analytic continuation will lead to periodic functions like $\cos\omega t_E$ and, in the literature, one usually restricts the range to $0<\omega t_E<2\pi$ often \textit{without} stating it as an assumption. This Euclidean range, when translated to Lorentzian sector, corresponds to limiting values of $\cosh\omega t=1$ at one limit and a rather strange value of $\cosh 2\pi$ at the other limit. The mere fact  the potential is periodic in $t_E$ does \textit{not} justify limiting to one cycle in the integration and this restriction of the range \textit{is always an extra assumption}. In a more general case, like the one we are studying, this assumption needs to be made explicit, as we have done.)
	
Let us now consider a general case  with an $F(\omega t)$ satisfying the above conditions and evaluate the relevant Euclidean action. We will assume that the field is in the $x$ direction and hence the vector potential also has only the $x$ component. The associated vector potential corresponds to, $A=-(E_{0}/\omega)F(\omega t)$ and hence the trajectory of the particle can be determined by solving the following differential equations
\begin{align}\label{EOM}
\frac{dx}{d\tau}=\frac{F(\omega t)}{\gamma};\qquad \frac{dt}{d\tau}=\sqrt{1+\frac{F^{2}(\omega t)}{\gamma ^{2}}}; \qquad \gamma \equiv\frac{m\omega}{qE_{0}}
\end{align}
The system is generically described by two independent, dimensionless,  constants $\gamma$ (which is a proxy for $\omega$) and $m^2/qE_0$. (As always we will assume that $qE_0$ is positive definite and stands for $|qE_0|$, though we will omit the modulus sign for notational simplicity.) For the computation of the relevant  action, we can restrict the trajectory to be along $x$ direction with initial momentum of the particle to be zero. Hence, using \ref{EOM} the action for a particle moving in the above electromagnetic field can be written in the following form,
\begin{align}
\mathcal{A}=-m\int d\tau +q\int A_{\alpha}(x)dx^{\alpha}=-\frac{m}{\gamma}\int dt~\sqrt{\gamma ^{2}+F^{2}}~.
\end{align}
 The  action in the Lorentzian sector can then be converted to the corresponding one in the Euclidean sector resulting in:
\begin{align}\label{int_general}
\mathcal{A}_{\rm E}&=i\mathcal{A}(t\rightarrow it_{E})=\frac{m}{\gamma}\int dt_{E} ~\sqrt{\gamma ^{2}+F^{2}(i\omega t_{\rm E})}\\
&=\frac{2m^{2}}{qE_{0}}\int _{-1}^{1} du ~\frac{\sqrt{1-u^{2}}}{F'(F^{-1}(\gamma u))}=\frac{2m^{2}}{qE_{0}}g(\gamma)~,
\end{align}
where $g(\gamma)$ corresponds to the value of the above integral, which is a function of the parameter $\gamma$. Given the Euclidean action, one can immediately obtain the associated particle number, which reads (see also \cite{Dumlu:2011cc}), 
\begin{align}
n_{\mathbf{k}}\approx\exp\left(-\mathcal{A}_{\rm E}\right)=\exp\left\{-\frac{2m^{2}}{qE_{0}}g(\gamma)\right\}~.
\end{align}
This result allows us to understand several features.

As we said before, the system is described by two constants $m^2/qE_0$ and $\gamma={m\omega}/{qE_{0}}$. \textit{If} we fix $\gamma$ and treat the result \textit{as a function of} $qE_0$, the result is \textit{always} non-analytic in $qE_0$ for the class of potentials we are studying. But if we fix $\omega$ (which is the natural parameter in the potential) and treat the result as a function of $qE_0$, the situation is more complicated (and interesting!). Since $\gamma$ has a dependence on $1/qE_0$, the overall dependence of the result on $qE_0$  depends sensitively on the nature of the function $g(\gamma)$. This, in turn, determines whether the result is analytic or non-analytic in the factor $qE_0$. For example, if, say, $g(\gamma)=1+\gamma$, then we will get a non-analytic factor in the exponent from $m^2/qE_0$; but if, say, $g(\gamma)=1/\gamma^2$, then the argument of exponent will pick up a factor $(1/qE_0)\times (qE_0)^2=qE_0$ and the result will be analytic in $qE_0$. If $g(\gamma)$ has two asymptotic 
limits (say, for small and large $\gamma$) varying between these two functional forms, then the result will change from being analytic to being non-analytic in $qE_0$ based on the value of $\gamma$. \textit{This is precisely how such transitions are interpreted in the literature.} 

This behaviour can be explicitly illustrated in the case of Sauter potential, for which the vector potential behaves as $-(E_{0}/\omega)\tanh (\omega t)$. Hence the function $F(\omega t)$ has the following expression: $F(\omega t)=\tanh (\omega t)$ and thus in the Euclidean regime it will become $i\tan (\omega t_{E})$. The above integration, presented in \ref{int_general} will now lead to:
\begin{align}
\mathcal{A}_{E}&=\frac{2m^{2}}{qE_{0}}\int _{-1}^{1} du\frac{\sqrt{1-u^2}}{\left[1+\gamma^2u^2\right]}=\frac{2m^{2}\pi}{qE_{0}}\frac{\left(\sqrt{1+\gamma ^{2}}-1\right)}{\gamma ^{2}}~.
\end{align}
which corresponds to the function $g(\gamma)$ given by: 
\begin{align}
g(\gamma)=\pi \frac{\left(\sqrt{1+\gamma ^{2}}-1\right)}{\gamma ^{2}}~.
\end{align}
One can now consider two limits, one corresponding to small values of $\gamma$, while the other one has to do with large values of $\gamma$. In the limit when the parameter $\gamma$ is small the function $g(\gamma)$ has the form
\begin{equation}
 g(\gamma)=\frac{\pi}{2}-\frac{\pi}{8}\gamma ^{2}+\mathcal{O}(\gamma ^{4}) ~.
\end{equation} 
Therefore, the number expectation will be given by 
\begin{align}
n_{\bf k}\approx \exp(-\mathcal{A}_{E})=\exp\left[-\frac{\pi m^2}{qE_0}\right]\left\{1+\frac{\pi m^{2}}{qE_{0}}\frac{\gamma ^{2}}{4}+\mathcal{O}(\gamma ^{4})\right\}~.
\end{align}
As evident when $qE_{0}\rightarrow 0$ the above expression is  non-analytic in the coupling constant (due to the $\exp(-1/x)$ behaviour of the leading term), which is consistent with the earlier results. On the other hand, in the large $\gamma$ limit we have:
\begin{align}
g(\gamma)\approx\frac{\pi}{\gamma}-\frac{\pi}{\gamma ^{2}}+ \mathcal{O}(\gamma ^{-3})
\end{align}
So, in this limit, the particle number will behave as,
\begin{align}
n_{\mathbf{k}}\approx\exp\left[-\frac{2m\pi}{\omega}\right]\left\{1+\frac{2\pi qE_{0}}{\omega ^{2}}+\mathcal{O}(\gamma ^{-3})\right\}~,
\end{align}
and thus will be analytic in the coupling constant (also see \cite{Dunne:2005sx}). Thus for small values of $\gamma$ the particle number is non-analytic, while for large values of $\gamma$ it turns out to be analytic. This explicitly depicts the transition of the particle number from being analytic to non-analytic as the parameter $\gamma$ transforms between large to small values.
In this case, the potential had a $\tanh\omega t$ function in the Lorentzian sector and thus becomes a periodic function in the Euclidean sector. Restricting the range of integration in the Euclidean sector seems justifiable in this context. 
\subsubsection{Polynomial potential}

In the case studied above, viz. the Sauter potential,  we worked with bounded function which could represent a realistic electric field --- in the sense that it vanishes at asymptotic times. There is another class of potentials, which --- though not as realistic --- are useful to address the following question: What happens to our results in \ref{nonperturbative} when several leading terms in the expansion vanish? For example, if $\tilde C_0=0$, the leading behaviour will be cubic in time; if the first two coefficients vanish the leading behaviour will scale as $t^5$ etc. This prompts us to study the case of a polynomial vector potential such that, $F(\omega t)=d_{1}\omega^{n} t^{n}$, where $n$ is assumed to be odd. (This is  unrealistic in the sense that the electric field is unbounded at large $t$ and should be treated as relevant in the intermediate times; but from this perspective even Schwinger electric field is also unrealistic, being constant for all time, and should be thought of as approximation to a field with asymptotic switching-off.) 

In this case the Euclidean action takes the following form,
\begin{align}
\mathcal{A}_{\rm E}=\frac{m}{\gamma}\int dt_{E} ~\sqrt{\gamma ^{2}-d_{1}^{2}\omega ^{2n} t_{\rm E}^{2n}}
\end{align}
The above integral can be converted to \ref{int_general} by the following substitution $d_{1}^{2}\omega ^{2n}t_{E}^{2n}=\gamma^{2}u^{2}$ and hence the function $g(\gamma)$ reads,
\begin{align}
g(\gamma)=\frac{1}{d_{1}^{1/n}\gamma ^{(n-1)/n}}\int _{-1}^{1}du ~u^{-(n-1)/n}\sqrt{1-u^{2}}
\end{align}
The above result essentially follows from the fact $\gamma du=d_{1}~n\omega ^{n}t_{E}^{n-1}dt_{E}$ and then one writes $t_{E}$ in terms of $u$. The terms outside the above integral are responsible for the analytic/non-analytic behaviour, since the integral itself is a pure number. Hence we obtain, 
\begin{align}
\mathcal{A}_{\rm E}=\frac{\sqrt{\pi}}{2}\frac{\Gamma\left(\frac{1}{2n}\right)}{\Gamma\left(\frac{3}{2}+\frac{1}{2n}\right)}\frac{2m\gamma^{1/n}}{\omega d_{1}^{1/n}}~.
\end{align}
Thus the number expectation will read,
\begin{align}
n_{\mathbf{k}}\approx\exp(-\mathcal{A}_{E})= \exp \left\{-\frac{\sqrt{\pi}}{2}\frac{\Gamma\left(\frac{1}{2n}\right)}{\Gamma\left(\frac{3}{2}+\frac{1}{2n}\right)}\frac{2m\gamma^{1/n}}{\omega d_{1}^{1/n}}\right\}
\end{align}
Thus for power law potential, the number expectation is indeed non-analytic, but goes only as $(qE_{0})^{-1/n}$, which is milder non-analyticity compared to $(qE_{0})^{-1}$. Thus the higher order powers will provide more and more sub-leading, non-analytic, contributions. 

As an example let us consider the case with just $\tilde C_0=0$ so that  the vector potential takes the following form, $F(\omega t)=d_{1}\omega ^{3}t^{3}$. Then the function  $g(\gamma)$ is given by:
\begin{align}
g(\gamma)=\frac{1}{d_{1}^{1/3}\gamma ^{2/3}}\int _{-1}^{1} du~u^{-2/3}~\sqrt{1-u^{2}}~.
\end{align}
Since the integral is independent of $\gamma$, all the dependence on $qE_{0}$, must come from the factors outside the integral. Thus the number expectation will read, 
\begin{align}
n_{\mathbf{k}}\approx\exp(-\mathcal{A}_{E})= \exp \left\{-\frac{\sqrt{\pi}}{2}\frac{\Gamma\left(\frac{1}{6}\right)}{\Gamma\left(\frac{5}{3}\right)}\frac{2m^{4/3}}{\omega ^{2/3}d_{1}^{1/3}(qE_{0})^{1/3}}\right\}
\end{align}
Thus the number expectation is indeed non-analytic, but only diverges for small $qE_{0}$ as $(qE_{0})^{-1/3}$ in the exponent, which is consistent with our previous analysis.  To summarise, in presence of the linear term, the Landau approach works and provides a non-analytic behaviour which goes as $\exp(-1/x)$. While for other powers of $t$, e.g., $F(\omega t)\sim \omega ^{3}t^{3}$, we will still have non-analyticity in the particle number, but  milder, varying only as $\exp(-1/x^{1/3})$ etc. 

Incidentally a similar scenario arises in the case of the following potential, $F(\omega t)=\tanh (\omega t)-\omega t$, which is essentially the Sauter potential minus the Schwinger potential. The Euclidean action can again be calculated and for small choices of $\gamma$ and the following expression for Euclidean action is obtained,
\begin{align}
\mathcal{A}_{\rm E}\simeq \frac{m}{\omega}\gamma ^{1/3}=\frac{m^{4/3}}{\omega ^{2/3}(qE_{0})^{1/3}}
\end{align}
Since to leading order, $\tanh(\omega t)-\omega t$ behaves as $\omega ^{3}t^{3}$, it is expected that the Euclidean action should match with the previous result for the $t^{3}$ potential and as evident, the Euclidean action does match with our previous considerations. 
	\section{Conclusion}
	
	The particle production in an external electric field serves as a toy model to understand various features of a generic QFT in an external classical background. An important feature of the particle production in the case of a \textit{constant} electric field (Schwinger effect) is that it is a non-perturbative phenomenon. This is reflected in the fact that, the mean number of particles produced with a particular momentum is a non-analytic function of the coupling constant and the field strength. However, when the external electric field is time-dependent, the mean number of particles produced could be either analytic or non-analytic in the coupling constant. A simple example which covers both the limits is provided by the Sauter type electric field in which case this behaviour is governed by a specific parameter in the potential. 
	
	Motivated by these facts, we have investigated a wider class of field configurations to contrast analytic versus non-analytic dependence in the coupling constant. We have 
considered two fairly general classes of electric fields, such that, for one class the particle production is analytic while for the other it is non-analytic in $qE_0$. We have shown that, for a class of electric fields, the non-analytic behaviour, at the leading order, is controlled by a single parameter which occurs in the theory. The situation becomes more complicated when we proceed beyond the leading order. But even in this context, one can make some progress and identify the general pattern if the electric field is such that Euclidean action techniques can be used. Using this, we show that the non-analytic behaviour becomes weaker when we proceed beyond the leading order.

The results and the techniques have implications in other contexts, like for example, particle production in an expanding universe, non-analyticity in presence of field superposition, resulting into enhanced particle production \cite{Schutzhold:2008pz}, which we hope to investigate in future publications. 
	\section*{Acknowledgements}
	 Research of S.C. is funded by the INSPIRE Faculty Fellowship (Reg. No. DST/INSPIRE/04/2018/000893) from Department of Science and Technology, Government of India. T.P's research is partially supported by the J.C.Bose Research Grant of DST, India.
\appendix
\labelformat{section}{Appendix #1} 
\labelformat{subsection}{Appendix #1} 
\section{A note on Landau's  Approach}\label{App_Add}
	
In this appendix, we will point out a few scenarios in which the Landau approach does not work. From which the contexts in which the Landau approach works should also be clear. To set the stage, let us consider first a situation, where it works. For that purpose, let us choose the following vector potential and hence the function $F(\omega t)$, such that, $F(\omega t)=C_{0}\omega t+C_{1}\omega ^{3}t^{3}$. In this case for small values of $\gamma$ the frequency becomes, 
\begin{align}
\omega_{\mathbf{k}}(t)\simeq \frac{m}{\gamma}\left(C_{0}\omega t+C_{1}\omega ^{3}t^{3}\right) +k_{x}+\frac{\gamma (\mathbf{k}_{\perp}^{2}+m^{2})}{2m}\frac{1}{\left(C_{0}\omega t+C_{1}\omega ^{3}t^{3}\right)}
\end{align}
where $k_{x}$ is the momentum of the particle along $x$ direction and $\mathbf{k}_{\perp}$ is the momentum along the directions orthogonal to $x$. Hence the frequency integral yields,
\begin{align}
\int dt~ \omega_{\mathbf{k}}(t)\simeq \frac{m}{\gamma}\left(\frac{C_{0}\omega}{2}t^{2}+\frac{C_{1}\omega ^{3}}{4}t^{4}\right) +k_{x}t+\frac{\gamma (\mathbf{k}_{\perp}^{2}+m^{2})}{2m\omega C_{0}}\log \left|\frac{\omega t}{\sqrt{C_{0}+C_{1}\omega ^{2}t^{2}}} \right|
\end{align}
Thus in the mode function we will have the following term, $(\omega t)^{i\gamma (\mathbf{k}_{\perp}^{2}+m^{2})/2m\omega C_{0}}$ and hence under $t\rightarrow e^{i\pi}t$, we obtain, the Bogoliubov coefficient to be, $\exp\{-\pi \gamma (\mathbf{k}_{\perp}^{2}+m^{2})/2m\omega C_{0}\}$. From which the number expectation will follow. Thus there is no place for the cubic part of the potential. To get non-trivial effects in the particle number from the cubic potential one must look at higher order WKB terms.

In the higher order WKB, we will have to evaluate the term involving integral of some function of frequency, since the only contribution, as we have seen in the previous situation, comes from logarithms. Thus to leading order in $\gamma$, the integral yields, 
\begin{align}
\frac{1}{8}\int \frac{\dot{\omega}_{\mathbf{k}}^2}{\omega_{k}^3}dt&=\frac{\gamma}{8m}\int dt~\frac{\dot{F}^{2}}{F^{3}}
=\frac{\omega \gamma}{16mC_{0}^{2}}\left[6C_{1}\log \left\{\frac{\omega t}{\sqrt{C_{0}+C_{1}\omega ^{2}t^{2}}}\right\}+\frac{-C_{0}^{3}+3C_{0}C_{1}^{2}\omega ^{4}t^{4}}{\omega ^{2}t^{2}\left(C_{0}+C_{1}\omega ^{2}t^{2}\right)} \right] 
\end{align}
Thus in the mode function following power-law behaviour will appear, $(\omega t)^{i3\omega \gamma C_{1}/8mC_{0}^{2}}$ and hence using Landau's approach we will have the following contribution to the Bogoliubov coefficient, $\exp\{-3\pi \omega \gamma C_{1}/8mC_{0}^{2}\}$. Thus the particle number for $\mathbf{k}=0$ calculated from the Landau approach, incorporating both the WKB terms, turns out to be
\begin{align}
n_0=\exp\left[-\frac{m^2\pi}{qE_0}\left(\frac{1}{C_0}+\frac{3C_1\omega ^{2}}{4C_0^2m^2}+\mathcal{O}(\gamma^2)\right)\right]
\end{align}
Thus here also we have non-analyticity in the coupling. However note that there is no way $C_{0}\rightarrow 0$ limit can be taken. Then one has to treat the case when $C_{0}=0$ in a separate manner by assuming the vector potential to behave as $C_{1}\omega ^{3}t^{3}$. In this case for small values of $\gamma$ the frequency becomes, 
\begin{align}
\omega_{\mathbf{k}}(t)\simeq \frac{m C_{1}}{\gamma}\omega ^{3}t^{3}+k_{x}+\frac{\gamma (\mathbf{k}_{\perp}^{2}+m^{2})}{2m}\frac{1}{C_{1}\omega ^{3}t^{3}}
\end{align}
and hence the integral of the frequency in the WKB approximation becomes,
\begin{align}
\int dt~ \omega_{\mathbf{k}}(t)\simeq \frac{m C_{1}}{4\gamma}\omega ^{3}t^{4} +k_{x}t-\frac{\gamma (\mathbf{k}_{\perp}^{2}+m^{2})}{4m\omega ^{3} C_{1}}\frac{1}{t^{2}}
\end{align}
As evident there are no logarithms present in the above computation and hence the Landau approach will provide trivial expressions for Bogoliubov coefficients. Thus Landau approach is \emph{not} suitable for computation of number expectation value associated with the $t^{3}$ potential.  

An identical scenario holds for the Sauter potential as well. For which if the large $\gamma$ expansion of the frequency $\omega _{\mathbf{k}}$ is taken, the integration appearing in the WKB approximation will lead to, 
\begin{align}
\int dt~\omega_{\mathbf{k}}(t)=\varepsilon t +\frac{k_{x}m}{\gamma \omega \varepsilon} \log \left|\cosh (\omega t)\right|+\mathcal{O}\left(\frac{1}{\gamma ^{2}}\right)
\end{align}
Even though $\log|\cosh(\omega t)|$ leads to a power-law term, it remains invariant under $t\rightarrow e^{i\pi}t$ and hence does not contribute in the Bogoliubov coefficient. Thus Landau approach \emph{cannot} provide the particle number in the case of Sauter potential in large $\gamma$ limit. Thus even though the Landau approach is important and useful in many occasions it does not work all the times and hence one must introduce some other technique, e.g., the Euclidean action technique to get an estimator of the particle number and hence non-analyticity. 

\providecommand{\href}[2]{#2}\begingroup\raggedright\endgroup

\end{document}